\documentclass[pra,aps,twocolumn,superscriptaddress]{revtex4}
\usepackage{amsfonts}
\usepackage{amsmath}
\usepackage{amssymb}
\usepackage{graphicx}
\usepackage{txfonts}

\begin{document}

\title{Data processing over single-port homodyne detection to realize
super-resolution and super-sensitivity}
\author{J. H. Xu}
\author{A. X. Chen}
\email{aixichen@zstu.edu.cn}
\affiliation{Key Laboratory of Optical Field Manipulation of Zhejiang Province and
Physics Department of Zhejiang Sci-Tech University, Hangzhou 310018, China}
\author{W. Yang}
\email{wenyang@csrc.ac.cn}
\affiliation{Beijing Computational Science Research Center, Beijing 100084, China}
\author{G. R. Jin}
\email{grjin@zstu.edu.cn}
\affiliation{Key Laboratory of Optical Field Manipulation of Zhejiang Province and
Physics Department of Zhejiang Sci-Tech University, Hangzhou 310018, China}

\begin{abstract}
Performing homodyne detection at one port of squeezed-state light
interferometer and then binarzing measurement data are important to achieve
super-resolving and super-sensitive phase measurements. Here we propose a
new data-processing technique by dividing the measurement quadrature into
three bins (equivalent to a multi-outcome measurement), which leads to a
higher improvement in the phase resolution and the phase sensitivity under
realistic experimental condition. Furthermore, we develop a new
phase-estimation protocol based on a combination of the inversion estimators
of each outcome and show that the estimator can saturate the Cram\'{e}r-Rao
lower bound, similar to asymptotically unbiased maximum likelihood estimator.
\end{abstract}

\maketitle


\section{Introduction}

Optimal measurement scheme followed by a proper data processing is important
to realize high-precision and high-resolution phase measurements~\cite{RMP,Ma,Dowling2008}. For the commonly used intensity measurement over
quasi-classical coherent states, the achievable phase sensitivity is subject
to the shot-noise limit (SNL) $\delta\theta\sim O(1/\sqrt{\bar{n}})$, where $\bar{n}$ is the number of particles of the input state. Furthermore, the
intensity measurement at the output-port of the coherent-state light
interferometer gives rise to an oscillatory interferometric signal $%
\propto\sin^{2}(\theta/2)$ or $\cos^{2}(\theta/2)$, which exhibits the
fringe resolution $\lambda/2$ determined by wavelength of the incident light
$\lambda$. This is often referred to the classical resolution limit of
interferometer, or the Rayleigh resolution criterion in optical imaging~\cite%
{Boto}. These two classical limits in the sensitivity and the resolution can
be surpassed with non-classical states of the light~\cite{Giovannetti2004,Durkin2007} such as the $N$-photon NOON state $(|N,
0\rangle_{a,b}+|0, N\rangle_{a,b})/\sqrt{2}$. This is a maximally entangled
state with all the particles being either in the mode $a$ or all in the mode
$b$, leading to the super-sensitivity $\delta\theta\sim O(1/N)$ and the
super-resolution $\lambda/(2N)$~\cite{Dowling2008,Boto,Giovannetti2004,Durkin2007,Gerry2003}. However, the NOON
states are difficult to prepare and are fragile to the loss-induced
decoherence~\cite{Braun,Dorner,Zhang2013a}.

Recently, several important progresses have been reported. The first one is the
achievement of super-resolution by feeding the interferometer with a
coherent laser, followed by coincidence photon counting~\cite{Resch}, parity
detection~\cite{Gao,Cohen}, and homodyne detection with a proper data
processing~\cite{Distante}. Specially, Distante \textit{et al}~\cite%
{Distante} detect the field quadrature at one port of coherent-state light
interferometer and then binarize the measurement data $p\in (-\infty,+\infty)$
into two bins $p\in\lbrack-a, a]$ and $p\notin\lbrack-a, a]$, which results
in a deterministic and robust super-resolution with classical states of the
light. The second progress is the recent theoretical proposal and
experimental demonstration~\cite{Schafermeier} that feeding a coherent state
and a squeezed vacuum state into the two input ports of the interferometer
followed by the same data processing over the single-port homodyne
detection, which can realize deterministic super-resolution and
super-sensitivity simultaneously with Gaussian states of light
and Gaussian measurements~\cite{Schafermeier}. This result may provide a powerful and efficient way to enhance the sensitivity of gravitational wave detectors~\cite{LIGO1,LIGO2} and that of correlation interferometry~\cite{Pradyumna}.

The data-processing method proposed by Refs.~\cite%
{Distante,Schafermeier} is equivalent to a binary-outcome measurement~\cite%
{Feng,Ghirardi,Jin}, where the outcome ``$0$" corresponds to $%
p\in\lbrack-a,a]$ and the outcome ``$\emptyset$" for $p\notin \lbrack-a, a]$%
. To infer an unknown phase shift, the simplest protocol of the phase
estimation has been used by inverting the averaged signal~\cite%
{Distante,Schafermeier}. The advantage of the inversion estimator is that it
has a relatively simple analytical expression and its sensitivity follows
the simple error-propagation formula~\cite{Distante,Schafermeier}. Moreover,
for any binary-outcome measurement, it has been shown that the inversion estimator
asymptotically saturates the Cram\'{e}r-Rao lower bound (CRB)~\cite%
{Feng,Ghirardi,Jin}. However, the binarization of measurement data and the
inversion estimator suffer from a serious drawback, i.e., they do not take
into account all the information from the measurement~\cite{YMK,Pezze}.
Consequently, they tend to degrade the achievable sensitivity significantly, e.g., at $\theta=0$, the sensitivity diverges~\cite{Distante,Schafermeier}, so the inversion estimator cannot infer the
true value of phase shift in the vicinity $\theta\sim0$.

In this paper, we propose a new strategy capable of further improving both
the resolution and the sensitivity using the experimental setup similar to
Schafermeier \textit{et al}~\cite{Schafermeier}. Our strategy consists of
two essential ingredients. The first one is to divide the measurement data
into three bins: $(-\infty,-a)$, $[-a,a]$, and $(a,\infty)$, corresponding
to three outcomes ``$-$", ``0", and ``$+$" , respectively. This is
equivalent to a three-outcome measurement and enjoys two advantages over the
previous binary-outcome case~\cite{Schafermeier}: (i) The divergence of
phase sensitivity at $\theta=0$ is removed, which is useful for estimating a
small phase shift; (ii) Higher improvement in the resolution and the
sensitivity is achievable under realistic experimental parameters. The
second ingredient is a composite estimator based on a linear combination of
the inversion estimators associated with each measurement outcome. This
estimator takes into account available information from all the measurement
outcomes of a general multi-outcome measurement, so it is capable of
saturating the CRB asymptotically. Therefore, this composite estimator
enjoys the good merits of the inversion estimator (i.e., the simplicity) and
the well-known maximum-likelihood estimator (i.e., unbiasedness and
asymptotic optimality in the sensitivity). In addition to the squeezed-state
light inteferometry, our estimation protocol may also be applicable to other
kinds of multi-outcome measurements.

\section{Single-port Homodyne detection without data-processing}

\begin{figure}[ptbh]
\begin{centering}
\includegraphics[width=0.9\columnwidth]{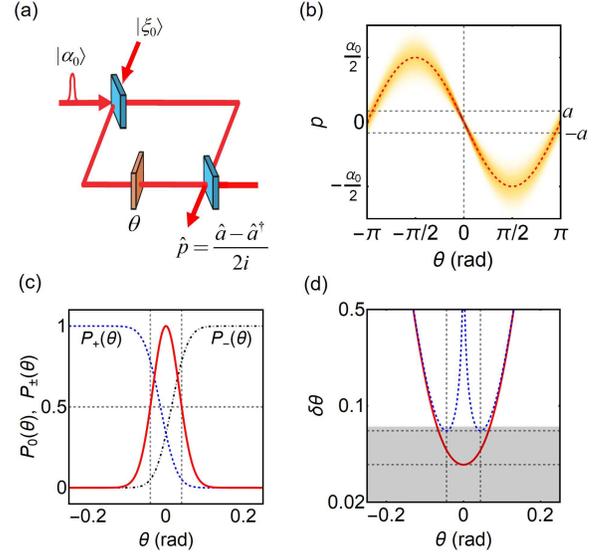}
\caption{(a) Homodyne detection (i.e., measuring the quadrature operator $\hat{p}$) at one port of the interferometer that fed by a coherent state $|\alpha_0\rangle$ and a squeezed vacuum $|\xi_0\rangle$. (b) Density plot of the probability $P(p|\theta)$ against the phase shift $\theta$ and the measurement quadrature $p$,
given by Eq.~(\ref{pro}). (c) Occurrence probabilities $P_{-}(\theta)$, $P_{0}(\theta)$, and $P_{+}(\theta)$ for detecting $p\in(-\infty, -a)$, $[-a, a]$, and $(a, \infty)$. This data-processing method is equivalent to a multi-outcome measurement. (d) The CRB of the phase sensitivity for the multi-outcome measurement $\delta\theta_{\mathrm{mul}}$ (red solid line), better than that of a binary-outcome measurement $\delta\theta_{\mathrm{bin}}$ (blue dashed line), where the measured data is divided into only two bins~\cite{Schafermeier}: $p\in[-a, a]$ and $p\notin[-a, a]$. Vertical lines in (c) and (d): the $\mathrm{FWHM}$ of the scaled $P_0(\theta)$ and the best
sensitivity of the binary-outcome measurement $\delta\theta_{\mathrm{bin},\min}=\delta\theta_{\mathrm{bin}}(\theta_{\min})$. The shaded area in (d): the region for the sensitivity better than the SNL $1/\sqrt{\bar{n}}$, where $\bar{n}=200$ (with $\alpha_0^2=199.3$) and the bin size $a=0.1$.}
\label{fig1}
\end{centering}
\end{figure}

As depicted by Fig.~\ref{fig1}(a), we consider the homodyne detection at one
port of the interferometer that fed by a coherent state $|\alpha _{0}\rangle
$ and a squeezed vacuum $|\xi _{0}\rangle $ (i.e., the so-called
squeezed-state interferometer)~\cite{Caves,Pezze2008}. To enlarge available
information about the phase shift $\theta $, the field amplitudes are chosen
as $\alpha _{0}\in \mathbb{R}$ and $\xi _{0}=-r\in \mathbb{R}$ (i.e., $\arg
\alpha _{0}=0$ and $\arg \xi _{0}=\pi $); See Refs.~\cite{LJing,LPan1,LPan2}
and also the Appendix. The total number of photons injected from the two
input ports is given by $\bar{n}=\alpha _{0}^{2}+\sinh ^{2}r$. Furthermore,
the Wigner function of the input state is given by~\cite{Gerrybook}
\begin{eqnarray}
W_{\mathrm{in}}(\alpha ,\beta ) &=&W_{|\alpha _{0}\rangle }(\alpha )W_{|\xi
_{0}\rangle }(\beta )  \notag \\
&=&\frac{2}{\pi }e^{-2\left[ \left( x_{a}-\alpha _{0}\right) ^{2}+p_{a}^{2}%
\right] }\cdot\frac{2\sqrt{\tilde{\mu}\tilde{\nu}}}{\pi }e^{-2\left( \tilde{%
\mu}x_{b}^{2}+\tilde{\nu}p_{b}^{2}\right) },  \label{Wigner_in}
\end{eqnarray}%
where $\alpha =x_{a}+ip_{a}$, $\beta =x_{b}+ip_{b}$, and
\begin{equation}
\tilde{\mu}=\varrho ^{2}e^{-2r},\text{ \ \ \ \ }\tilde{\nu}=e^{2r},
\label{uv}
\end{equation}%
with $\varrho $ ($\leq 1$) and $e^{-r}$ describing the purity and the
squeeze parameter of $|\xi _{0}\rangle $. The Wigner function of the output
state takes the same form with the input state $W_{\mathrm{out}}(\alpha
,\beta ;\theta )=W_{\mathrm{in}}(\tilde{\alpha}_{\theta },\tilde{\beta}%
_{\theta })$~\cite{Seshadreesan2,Tan,Wang}, where the variables $(\alpha
,\beta )$ have been replaced by $(\tilde{\alpha}_{\theta },\tilde{\beta}%
_{\theta })$; see the Appendix. Integrating the Wigner function over $%
\{x_{a} $, $x_{b}$, $p_{b}\}$, we obtain the conditional probability for
detecting a measurement quadrature $p\in (-\infty, \infty)$,
\begin{equation}
P(p|\theta )=\sqrt{\frac{2}{\pi \eta _{\theta }}}\exp \left[ -\frac{2}{\eta
_{\theta }}\left( p+\frac{\alpha _{0}}{2}\sin \theta \right) ^{2}\right] ,
\label{pro}
\end{equation}%
where, for brevity, we omit the subscript \textquotedblleft $a$" in the
quadrature $p_{a}$, and introduce
\begin{equation}
\eta _{\theta }=\frac{\tilde{\mu}+\tilde{\nu}+2\tilde{\mu}\tilde{\nu}-2%
\tilde{\mu}(\tilde{\nu}-1)\cos \theta +(\tilde{\mu}-\tilde{\nu})\cos
^{2}\theta }{4\tilde{\mu}\tilde{\nu}}.  \label{eta}
\end{equation}%
Note that Eq.~(\ref{pro}) holds for the homodyne detection at one port of
the interferometer fed by the input $|\alpha _{0}\rangle \otimes |\xi
\rangle $. Here $|\xi \rangle $ could be arbitrary gaussian state of light,
with $\tilde{\mu}$ and $\tilde{\nu}$ to be determined by $|\xi \rangle $. As
the simplest case, the coherent-state input $|\alpha _{0}\rangle \otimes
|0\rangle $ corresponds to $\tilde{\mu}=\tilde{\nu}=\varrho=1$ and hence $%
\eta_{\theta}=1$, in agreement with our previous result~\cite{Feng}.

In Fig.~\ref{fig1}(b), we show density plot of $P(p|\theta)$ against the
phase shift $\theta $ and the measurement quadrature $p$, where the red
dashed line is given by $p=-\alpha_{0}\sin(\theta)/2$. This equation takes
the same form with that of the signal
\begin{equation}
\langle \hat{p}(\theta )\rangle =\int_{-\infty }^{\infty }P(p|\theta )pdp=-%
\frac{\alpha _{0}}{2}\sin \theta,
\end{equation}%
which shows the full width at half maximum ($\mathrm{FWHM}$) $=2\pi/3$, and hence the Rayleigh limit in
fringe resolution~\cite{Distante,Schafermeier}.

According to Refs.~\cite{Helstrom,Braunstein1,Braunstein2,Paris2009}, the ultimate phase estimation precision is determined by the CFI:
\begin{eqnarray}
\mathcal{F}(\theta ) &=&\int_{-\infty }^{\infty }\frac{[P^{\prime }(p|\theta
)]^{2}}{P(p|\theta )}dp  \notag \\
&=&\frac{(\alpha _{0}\cos \theta )^{2}}{\eta _{\theta }}+\frac{\left[ \tilde{%
\mu}-\tilde{\mu}\tilde{\nu}+(\tilde{\mu}-\tilde{\nu})\cos \theta \right]
^{2}\sin ^{2}\theta }{8(\tilde{\mu}\tilde{\nu}\eta _{\theta })^{2}},
\end{eqnarray}%
where $P^{\prime}\equiv\partial P/\partial\theta$. When the coherent-state
component dominates over the squeezed vacuum, maximum of the CFI occurs at $%
\theta=0$, i.e., $\mathcal{F}(0)=\alpha_{0}^{2}/\eta_{0}=\tilde{\nu}%
\alpha_{0}^{2}\simeq e^{2r}\bar{n}$, which yields a sub-shot-noise
sensitivity:
\begin{equation}
\delta \theta _{\mathrm{CRB},\mathrm{\min }}=\frac{1}{\sqrt{\mathcal{F}(0)}}%
\simeq \frac{e^{-r}}{\sqrt{\bar{n}}}.  \label{sub_SNL}
\end{equation}%
This is the best sensitivity attained from the single-port homodyne
detection in the limit
$\alpha_{0}^{2}\gg\sinh^{2}r$, coincident with the intensity-difference measurement~\cite{Caves,Pezze2008}.

\section{Binary-outcome homodyne detection}

To improve the resolution, one can separate the measured data into two bins~%
\cite{Distante}: $p\in \lbrack -a,a]$ as an outcome, denoted by
``$0$", and $p\notin \lbrack -a,a]$ as an another outcome
``$\emptyset $", with the bin size $2a$. Using Eq.~(\ref{pro}%
), it is easy to obtain the conditional probabilities of the outcomes,
\begin{equation}
P_{0}(\theta )=\int_{-a}^{+a}dpP(p|\theta )=\frac{1}{2}\mathrm{Erf}\left[
g_{-}(\theta ),g_{+}(\theta )\right] ,  \label{Pplus_cs}
\end{equation}%
and hence $P_{\emptyset }(\theta )=1-P_{0}(\theta )$. Here, $\mathrm{Erf}[
x, y] =\mathrm{erf}(y)-\mathrm{erf}(x)$ denotes a generalized error
function, and
\begin{equation}
g_{\pm }(\theta )=\sqrt{\frac{2}{\eta _{\theta }}}\left( \frac{\alpha _{0}}{2%
}\sin \theta \pm a\right) ,  \label{gphi}
\end{equation}%
with $\eta_{\theta}$ being defined in Eq.~(\ref{eta}). The above
data-processing method is equivalent to a binary-outcome measurement~\cite%
{Wang}, with the observable $\hat{\Pi}=\mu _{0}\hat{\Pi}_{0}+\mu _{\emptyset
}\hat{\Pi}_{\emptyset }$, where $\hat{\Pi}_{0}=\int_{-a}^{+a}|p\rangle
\langle p|dp $ and $\hat{\Pi}_{\emptyset }=\hat{1}-\hat{\Pi}_{0}$.
Obviously, the output signal is given by
\begin{equation}
\langle \hat{\Pi}(\theta )\rangle =\mu _{0}P_{0}(\theta )+\mu _{\emptyset
}P_{\emptyset }(\theta ),  \label{sinal2}
\end{equation}%
where we have used the relation $\langle \hat{\Pi}_{k}(\theta )\rangle
=P_{k}(\theta )$ for $k=0$ and $\emptyset$. Following Schafermeier \textit{%
et al}~\cite{Schafermeier}, in Fig.~\ref{fig1}(c), we choose the eigenvalues
$\mu_{\emptyset }=0$ and $\mu_{0}=1/\mathrm{erf}(\sqrt{2}ae^{r})$ to show
the signal as a function of $\theta$ (see the red solid line), which shows $%
\langle\hat{\Pi}(0)\rangle=1$. This treatment is useful to determine the $\mathrm{FWHM}$ of the signal and hence the resolution, as depicted by the vertical lines of Fig.~\ref{fig1}(c).

\begin{figure}[ptbh]
\begin{centering}
\includegraphics[width=0.9\columnwidth]{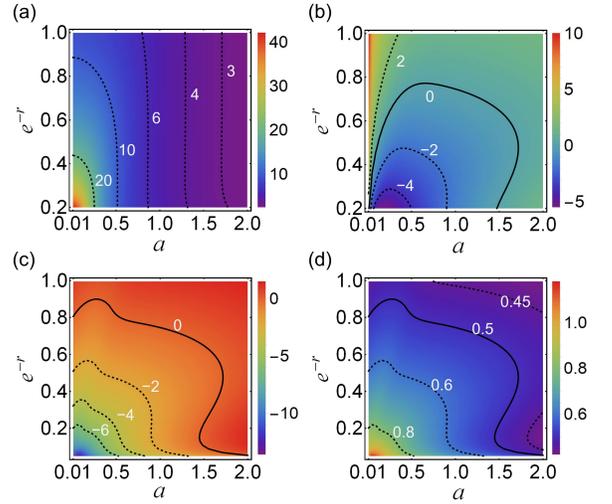}
\caption{For the purity of the squeezed vacuum $\varrho=0.5$ and given number of photons $\bar{n}=\alpha_0^2+\sinh^2r=100$, density plot of the improvement factor in the resolution $\frac{2\pi/3}{\mathrm{FWHM}}$ (a) against the bin size $a$ and the squeeze parameter $e^{-r}$, and that of the best sensitivities (in units of $\mathrm{dB}$), obtained from the binary-outcome measurement $\delta\theta_{\mathrm{bin},\min}$ (b) and the multi-outcome measurement $\delta\theta_{\mathrm{mul},\min}$ (c). (d) The scaling of the sensitivity $\frac{-\log\delta\theta_{\mathrm{mul},\min}}{\log\bar{n}}$ ($=0.5$ for the shot-noise limit, and $1$ for the Heisenberg limit). The solid lines in (b)-(d): the shot-noise limit.}
\label{fig2}
\end{centering}
\end{figure}

In Fig.~\ref{fig2}(a), we show numerical results of the $\mathrm{FWHM}$ as
functions of the bin size $a $ and the squeezing parameter $e^{-r}$. Similar
to Ref.~\cite{Schafermeier}, one can note that the improvement of the $%
\mathrm{FWHM}$ compared to the Rayleigh criterion $2\pi/3$ (i.e., the ratio $%
\frac{2\pi /3}{\mathrm{FWHM}}$) increases as $a\rightarrow 0$ and $%
r\rightarrow \infty$. For a given and finite number of photons $\bar{n}$,
this means that a better resolution beyond the Rayleigh criterion (i.e., the
super-resolution) can be obtained when $a$, $\alpha _{0}\rightarrow 0$.

Independent on $\mu_{0}$ and $\mu_{\emptyset}$, the phase sensitivity of
the binary-outcome measurement is given by
\begin{equation}
\delta \theta _{\mathrm{bin}}=\frac{\Delta \hat{\Pi}}{|\partial \langle \hat{%
\Pi}(\theta )\rangle /\partial \theta |}=\frac{\sqrt{P_{0}(\theta
)P_{\emptyset }(\theta )}}{\left\vert P_{0}^{\prime }(\theta )\right\vert },
\end{equation}%
where $\Delta \hat{\Pi}=\sqrt{\langle \hat{\Pi}^{2}\rangle -\langle \hat{\Pi}%
\rangle ^{2}}$ and $P_{0}^{\prime }=\partial P_{0}/\partial \theta $. On the
other hand, the CFI of this binary-outcome measurement is given by~\cite%
{Feng}
\begin{equation}
\mathcal{F}_{\mathrm{bin}}(\theta )=\sum_{k=0,\emptyset }\frac{\left[
P_{k}^{\prime }(\theta )\right] ^{2}}{P_{k}(\theta )}=\frac{1}{(\delta
\theta_{\mathrm{bin}})^{2}},
\end{equation}%
where, in the last step, we have used the normalization relation $%
P_{0}(\theta )+P_{\emptyset }(\theta )=1$. The above results indicate that
the phase uncertainty predicted by the error-propagation $\delta \theta _{%
\mathrm{bin}}$ always saturates the CRB $1/\sqrt{\mathcal{F}_{\mathrm{bin}%
}(\theta )}$, which holds for any binary-outcome measurement~\cite%
{Feng,Ghirardi,Jin}. As illustrated by the blue dashed line of Fig.~\ref%
{fig1}(d), one can see that the sensitivity reaches its maximum at the
optimal working point $\theta _{\min }$ (the vertical lines) and the best
sensitivity $\delta \theta _{\mathrm{bin,\min }}\equiv \delta \theta _{%
\mathrm{bin}}(\theta _{\min })$ can beat the \textrm{SNL} ($=1/\sqrt{\bar{n}}
$).

Similar to Ref.~\cite{Schafermeier}, in Fig.~\ref{fig2}(b), we show the
improvement in the sensitivity $\delta \theta _{\mathrm{bin,\min }}/\mathrm{%
SNL}$ as functions of the bin size $a$ and the squeezing parameter $e^{-r}$.
For a given $\bar{n}=100$, the best sensitivity can reach $4\mathrm{dB}$
when $a=0.5$ and $e^{-r}=0.2$ (i.e., $\sinh ^{2}r/\bar{n}\approx 0.06$).
From the squares of Fig.~\ref{fig3}, one can also find that the $\mathrm{FWHM%
}$ scales as $(2\pi /3)/\sqrt{\bar{n}}$ and the best sensitivity $\delta
\theta _{\mathrm{bin,\min }}\sim 0.75/\bar{n}^{0.54}$, with the scaling
better than the $\mathrm{SNL}$ (i.e., the super-sensitivity). Specially, a $%
22$-fold improvement in the phase resolution and a $1.7$-fold improvement in
the sensitivity can be obtained with $a=0.5$, $\alpha_{0}^{2}=427$, and $%
\sinh^{2}r=0.687$ (i.e., $e^{-r}=0.47$)~\cite{Schafermeier}.

Normally, the data processing over the measurement quadrature $p\in(-\infty, \infty)$ can increase the resolution, at the cost of reduced phase sensitivity. In this sense, the ultimate phase sensitivity obtained from the single-port
homodyne measurement without any data-processing (i.e., $\delta\theta_{\mathrm{CRB},\mathrm{\min}}$) is the best sensitivity of the
binary-outcome measurement in the limit $a\rightarrow\infty$~\cite{Schafermeier}. From Fig.~\ref{fig3}, one can see $\delta \theta _{\mathrm{bin,\min }}>\delta \theta _{\mathrm{CRB}\text{,}\mathrm{\min}}$ (the thick solid line). More
importantly, $\delta\theta_{\mathrm{bin}}$ diverges at $\theta =0$ and
therefore \emph{no} phase information can be inferred for a small phase
shift $\theta \sim 0$. To avoid this problem, we present a new
data-processing technique (equivalent to a multi-outcome measurement), based
upon the experimental setup similar to Schafermeier \textit{et al}~\cite%
{Schafermeier}.

\section{Multi-outcome homodyne detection}

We now consider a new data-processing method by treating the measurement
quadrature $p\in (a, \infty)$ as an outcome, denoted hereinafter by ``$+$",
and similarly $p\in (-\infty, -a)$ as an outcome ``$-$". The conditional
probabilities for detecting ``$\pm $" are given by%
\begin{eqnarray}
P_{+}(\theta ) &=&\int_{a}^{\infty }dpP(p|\theta )=\frac{1-\mathrm{erf}%
[g_{+}(\theta )]}{2},  \label{proPlus} \\
P_{-}(\theta ) &=&\int_{-\infty }^{-a}dpP(p|\theta )=\frac{1+\mathrm{erf}%
[g_{-}(\theta )]}{2},  \label{proMinus}
\end{eqnarray}%
which obey the normalization condition $P_{+}(\theta )+P_{0}(\theta
)+P_{-}(\theta )=1$ and $g_{\pm }(\theta )$ have been defined in Eq.~(\ref%
{gphi}). This is indeed a multi-outcome measurement with the observable $%
\hat{\Pi}=\sum_{k}\mu _{k}\hat{\Pi}_{k}$~\cite{Wang}, defined by the
projections $\hat{\Pi}_{+}=\int_{a}^{\infty }|p\rangle \langle p|dp$, $\hat{%
\Pi}_{-}=\int_{-\infty }^{-a}|p\rangle \langle p|dp$, and $\hat{\Pi}_{0}$.
In Fig.~\ref{fig1}(c), we show $P_0$ and $P_{\pm}$ as functions of $\theta$,
for $\bar{n}=200$, $a=0.1$, and the purity $\varrho=1$. Hereinafter, we choose a relatively small value of $a$ than that of Ref.~\cite{Schafermeier} to obtain a better resolution and an enhanced sensitivity [see below Figs.~\ref{fig2}(c) and (d)].

For a general multi-outcome measurement, the averaged signal can be obtained
by taking expectation value of $\hat{\Pi}$ with respect to a phase-encoded
state $\hat{\rho}(\theta)$, namely%
\begin{equation}
\langle \hat{\Pi}(\theta )\rangle =\sum_{k}\mu _{k}P_{k}(\theta )\approx
\sum_{k}\mu _{k}\frac{{\mathcal{N}}_{k}}{{\mathcal{N}}},  \label{signal}
\end{equation}%
where $\mu_{k}$ and $P_{k}(\theta)=\langle \hat{\Pi}_{k}\rangle =\mathrm{Tr}[%
\hat{\rho}(\theta )\hat{\Pi}_{k}]$ denote the eigenvalue and the conditional
probability associated with the $k$th outcome. With ${\mathcal{N}}$
independent measurements, one records the occurrence number of each outcome $%
{\mathcal{N}}_{k}$ at given $\theta \in (-\pi ,\pi )$. As ${\mathcal{N}}\gg
1 $, the conditional probabilities can be measured by the occurrence
frequencies, due to $P_{k}(\theta )\approx {\mathcal{N}}_{k}/{\mathcal{N}}$.
For the multi-outcome homodyne measurement, we numerical simulate $%
P_{0}(\theta)$ and $P_{\pm}(\theta)$ using $M$ replicas of ${\mathcal{N}}$
random numbers~\cite{Wang}. As illustrated by the solid circles of Fig.~\ref%
{fig4}(a) and (b), one can note that statistical average of the occurrence
frequencies ${\mathcal{N}}_{0}/{\mathcal{N}}$ and ${\mathcal{N}}_{\pm }/{%
\mathcal{N}}$, fitted as $P_{0}^{\mathrm{(fit)}}(\theta ) $ and $P_{\pm }^{%
\mathrm{(fit)}}(\theta)$, show good agreement with their analytical results.

Once all phase-dependent $\{P_{k}(\theta)\}$ and hence $\langle\hat{\Pi}%
(\theta )\rangle$ are known, one can infer $\theta $ via the inversion
estimator $\theta_{\mathrm{inv}}=g^{-1}(\sum_{k}\mu _{k}{\mathcal{N}}_{k}/{%
\mathcal{N}})$, where $g^{-1}$ denotes the inverse function of $g(\theta
)=\langle \hat{\Pi}(\theta )\rangle $. This protocol of phase estimation is
commonly used in experiments, since its performance simply follows the
error-propagation formula. However, the inversion estimator based on the
averaged signal does not take into account all of the available information,
especially the fluctuations in the measurement observable at the output
ports~\cite{YMK}. To improve the phase information, one can adopt
data-processing techniques such as maximal likelihood estimation or Bayesian
estimation~\cite{Pezze}, which saturates the CRB~\cite%
{Helstrom,Braunstein1,Braunstein2,Paris2009}:%
\begin{equation}
\Delta \theta _{\mathrm{mul}}=\frac{1}{\sqrt{{\mathcal{N}}\mathcal{F}_{%
\mathrm{mul}}(\theta )}},  \label{CRB}
\end{equation}%
where $\mathcal{F}_{\mathrm{mul}}(\theta )=\sum_{k}f_{k}(\theta )$, being a
sum of the CFI of each outcome, with
\begin{equation}
f_{k}(\theta )=\frac{1}{P_{k}(\theta )}\left[ \frac{\partial P_{k}(\theta )}{%
\partial \theta }\right] ^{2}.  \label{cfi}
\end{equation}%
The phase-dependent $\{P_{k}(\theta)\}$ and hence $\{f_{k}(\theta)\}$ can be
obtained in principle, at least, from the interferometric calibration, where
the value of $\theta$ is known and tunable.

In Fig.~\ref{fig1}(d), we show the sensitivity per measurement $\delta
\theta _{\mathrm{mul}}\equiv \sqrt{\mathcal{N}}\Delta \theta _{\mathrm{mul}}$
as a function of $\theta $ (the red line). The best sensitivity occurs at $%
\theta =0$ and hence $\delta \theta _{\mathrm{mul}\text{,}\min }\equiv 1/%
\sqrt{\mathcal{F}_{\mathrm{mul}}(0)}$. The improvement of $\delta \theta _{%
\mathrm{mul}\text{,}\min }$ compared with the $\mathrm{SNL}$ is depicted in
Fig.~\ref{fig2}(c), which shows larger quantum-enhancement region than that
of $\delta \theta _{\mathrm{bin}\text{,}\min}$. In
Fig.~\ref{fig2}(d), we show the
scaling of the best sensitivity $\frac{-\log\delta\theta_{\mathrm{mul}\text{,%
}\min }}{\log \bar{n}}$ against the bin size $a$ and the squeezing parameter
$e^{-r}$, where the solid line implies the $\mathrm{SNL}$. For a given $\bar{n}\gg1$, one can find that the scaling can even reach the Heisenberg limit as $\alpha _{0}$, $a\rightarrow 0$.

\begin{figure}[phtb]
\begin{centering}
\includegraphics[width=1\columnwidth]{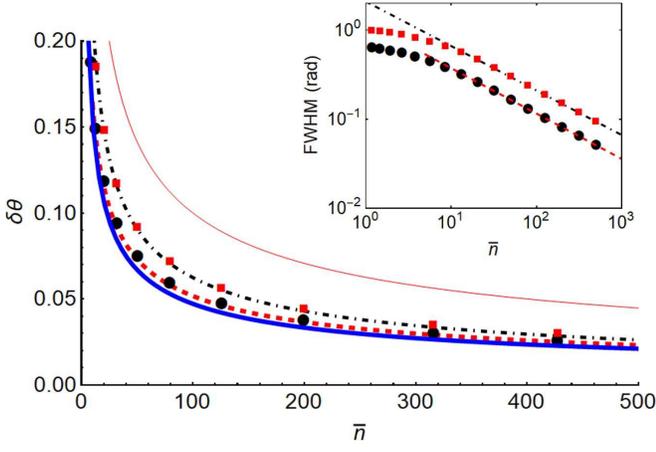}
\caption{For given $\sinh^2r=0.687$ (i.e., $e^{-r}=0.47$) and the purity $\varrho=0.58$, the best sensitivity as a function of $\bar{n}$ for the multi-outcome measurement with $a=0.1$ (solid circles), and that of the binary-outcome measurement with $a=0.5$ (red squares) and $a\rightarrow\infty$ (blue thick line, given by Eq.~(\ref{sub_SNL})). Red thin line: the SNL. Dot-dashed line: $\delta \theta _{\mathrm{bin,\min}}\sim0.75/\bar{n}^{0.54}$; Red dashed line: $\delta \theta _{\mathrm{mul,\min}}\sim1.1e^{-r}/\sqrt{\bar{n}}$. Inset: the FWHM of the scaled $P_0(\theta)$ as a function of $\bar{n}$ for $a=0.5$ (red squares) and $0.1$ (solid circles). The dot-dashed line in the inset: $(2\pi/3)/\sqrt{\bar{n}}$; The red dashed line: $1.21/\bar{n}^{0.51}$.}
\label{fig3}
\end{centering}
\end{figure}

In Fig.~\ref{fig3}, we show the scaling of $\delta\theta
_{\mathrm{mul}\text{,}\min}$ and compare it with $\delta\theta
_{\mathrm{bin}\text{,}\min}$, using the parameters $\sinh^{2}r=0.687$ and $\varrho=0.58$. To optimize the performance, we choose the bin size $a=0.1$ for the multi-outcome measurement; While for the binary-outcome case, we take $a=0.5$~\cite{Schafermeier}. One can find that numerical results of $\delta\theta_{\mathrm{mul}\text{,}\min}$ (the solid circles) can be well fitted as $1.1e^{-r}/\sqrt{\bar{n}}$, better than that of $\delta\theta_{\mathrm{bin}\text{,}\min}$ (the squares). This result almost approaches the best sensitivity of the single-port
homodyne measurement without any data-processing (the thick line). From the inset, one can
also note that the signal becomes further narrowing in a comparison with
that of Ref.~\cite{Schafermeier}. For instance, a $38$-fold improvement in
the resolution and a $1.9$-fold improvement in the sensitivity is achievable
with the realistic experimental parameters~\cite{Schafermeier}: $a=0.1$, $\alpha_{0}^{2}=427$, and $\sinh^{2}r=0.687$.

\begin{figure}[ptbh]
\begin{centering}
\includegraphics[width=1\columnwidth]{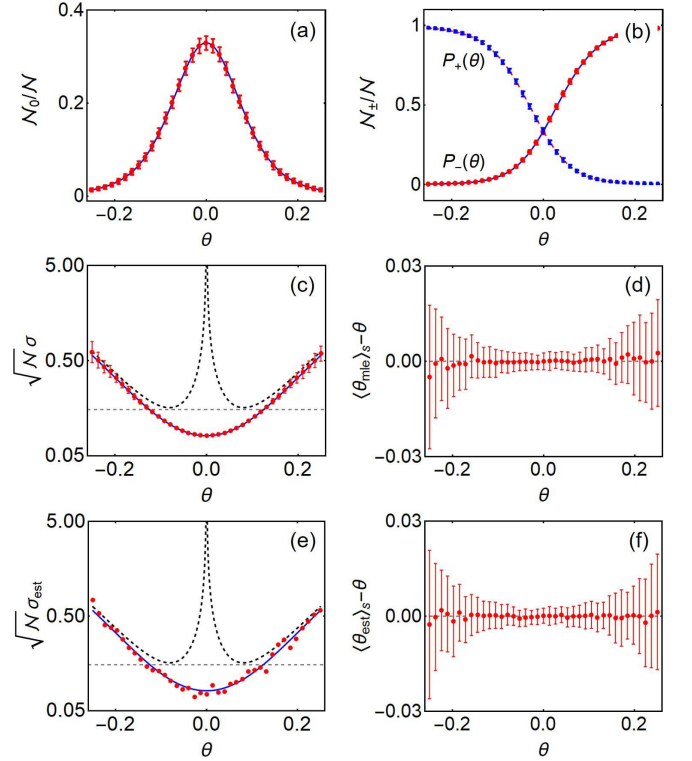}
\caption{With $M=100$ replicas of ${\mathcal{N}}=1000$ random numbers at each given $\theta\in(-1/4, 1/4)$, (a) and (b) statistical average of the
occurrence frequencies ${\mathcal{N}}_{0}/{\mathcal{N}}$ and ${\mathcal{N}}%
_{\pm }/{\mathcal{N}}$ (solid circles) and their standard derivations (the bars), following $P_{0 }(\theta)$ and $P_{\pm}(\theta)$. (c) and (e) The averaged phase uncertainty per measurement of $\theta_{\mathrm{mle}}$ and $\theta_{\mathrm{est}}$, following $\delta\theta_{\mathrm{mul}}$ (blue solid lines) and surpassing the phase sensitivity of the binary-outcome measurement $\delta\theta_{\mathrm{bin}}$ (dashed lines). (d) and (f) The bias of $\theta_{\mathrm{mle}}$ and $\theta_{\mathrm{est}}$ (solid circles), smaller than their standard derivations (the bars), indicating unbiasedness of them. The horizontal lines in (c) and (e): the $\mathrm{SNL}$. Other parameters: $\varrho=0.58$, $a=0.1$, $\alpha_0^2=42$, and $\sinh^2r=0.687$ ($e^{-r}=0.47$).}
\label{fig4}
\end{centering}
\end{figure}

To saturate the CRB, we adopt two estimation protocols based on the
single-port homodyne detection in the squeezed-state interferometer. The
first one is maximum-likelihood estimation. It is well known that the MLE is
unbiased and can saturate the CRB when ${\mathcal{N}}\gg 1$ (see e.g. Ref.~%
\cite{Helstrom}). Numerically, the estimator $\theta _{\mathrm{mle}}$ can be
determined by maximizing the likelihood function (i.e., a multinomial
distribution):
\begin{equation}
\mathcal{P}(\theta |\{{\mathcal{N}}_{k}\})={\mathcal{N}}!\prod_{k}\frac{1}{{%
\mathcal{N}}_{k}!}\left[ P_{k}^{\mathrm{(fit)}}(\theta )\right] ^{{\mathcal{N%
}}_{k}},  \label{likelihood}
\end{equation}%
where ${\mathcal{N}}_{k}={\mathcal{N}}_{k}(\theta _{0})$ denotes the
occurrence number of each outcome at a given true value of phase shift $%
\theta _{0}$, and $P_{k}^{\mathrm{(fit)}}(\theta )$ is a fit of the averaged
occurrence frequency. To speed up numerical simulations, we directly use the
analytical results of $P_{k}(\theta)$. For large enough $\mathcal{N}$, the
phase distribution can be well approximated by a Gaussian~\cite{Jin}:
\begin{equation}
\mathcal{P}(\theta |\{{\mathcal{N}}_{k}\})\varpropto \exp \left[ -\frac{%
(\theta -\theta _{\mathrm{mle}})^{2}}{2\sigma ^{2}}\right] ,
\label{Gaussian}
\end{equation}%
where $\sigma$ is $68.3\%$ confidence interval of the Gaussian around $%
\theta _{\mathrm{mle}}$, determined by
\begin{equation}
\sigma \approx \sqrt{\frac{1}{\left\vert \partial ^{2}\mathcal{P}(\theta |\{{%
\mathcal{N}}_{k}\})/\partial \theta ^{2}\right\vert }}.  \label{sigma}
\end{equation}%
In Fig.~\ref{fig4}(c), we plot the averaged phase uncertainty per
measurement $\sqrt{{\mathcal{N}}}\sigma$ (see the circles) and its standard
derivation (the bars) for each given $\theta _{0}$, using $M$ replicas of ${%
\mathcal{N}}$ random numbers. One can find that the circles follows the blue
solid line (i.e., $\delta \theta _{\mathrm{mul}}$). Furthermore, from Fig.~%
\ref{fig4}(d), one can find that standard derivation of $\theta_{\mathrm{mle}%
}$ (the bars) is larger than averaged value of the error $(\theta_{\mathrm{%
mle}}-\theta_{0})$, indicating that $\theta_{\mathrm{mle}}$ is unbiased~\cite%
{Pezze}.

A new phase-estimation protocol can be obtained from a convex
combination of the CFI of each outcome $f_{k}(\theta)$. First, we define
the inversion estimator of each outcome $\theta _{\mathrm{inv},k}=P_{k}^{-1}(%
{\mathcal{N}}_{k}/{\mathcal{N}})$ by inverting the equation $P_{k}(\theta )={%
\mathcal{N}}_{k}/{\mathcal{N}}$. Next, we construct a composite phase
estimator with the weight determined by $f_{k}(\theta )$,
\begin{equation}
\theta _{\mathrm{est}}=\sum_{k}c_{k}\theta _{\mathrm{inv},k},\text{ \ \ \ }%
c_{k}=\frac{f_{k}\left( \theta _{\mathrm{inv},k}\right) }{%
\sum_{k}f_{k}\left( \theta _{\mathrm{inv},k}\right) },  \label{estimator}
\end{equation}%
where $f_{k}(\theta )$ has been defined by Eq.~(\ref{cfi}), with $k=0$, $\pm
$ for the multi-outcome homodyne measurement. Obviously, this result is
physically intuitive. For example, if the CFI of the outcome $k=0$ dominates
over that of the others (so that $\theta_{\mathrm{inv},0}$ is much
more reliable than $\theta _{\mathrm{inv},\pm}$), then the above equation
reduces to $\theta _{\mathrm{est}}\approx \theta _{\mathrm{inv},0}$.
Furthermore, this estimator enjoys the good merits of the inversion
estimator (i.e., the simplicity) and the well-known maximum-likelihood
estimator (i.e., unbiasedness and asymptotic optimality in the sensitivity).
In Fig.~\ref{fig4}(e) and (f), we numerically obtain the estimators $%
\{\theta _{\mathrm{est}}^{(1)},\theta _{\mathrm{est}}^{(2)},\cdots ,\theta _{%
\mathrm{est}}^{(M)}\}$ using $M$ replicas of ${\mathcal{N}}$ random numbers
at each given $\theta _{0}$. Unlike the MLE, the performance of $\theta _{%
\mathrm{est}}$ is simply determined by the root-mean-square fluctuation
\begin{equation}
\sigma _{\mathrm{est}}=\sqrt{\langle (\theta _{\mathrm{est}}^{(i)}-\theta
_{0})^{2}\rangle _{s}},  \label{RMSE}
\end{equation}
where $\langle (\cdots)\rangle _{s}\equiv
\sum_{i=1}^{M}(\cdots)/M$ denotes the statistical average. As shown in Fig.~%
\ref{fig4}(e) and (f), one can find that the averaged phase uncertainty per
measurement $\sqrt{{\mathcal{N}}}\sigma _{\mathrm{est}}$ almost follows the
CRB $\delta \theta _{\mathrm{mul}}$ and the bias $\langle \theta _{\mathrm{%
est}}\rangle _{s}-\theta _{0}$ is almost vanishing, similar to the MLE.

It should be mentioned that the dashed lines in Fig.~\ref{fig4}(c) and (e) show the sensitivity of the binary-outcome scheme $\delta\theta
_{\mathrm{bin}}$, which can beat the $\mathrm{SNL}$ if one takes $a=0.5$ (see Ref.~\cite{Schafermeier}). Based on Eq.~(\ref{signal}), one can also investigate the performance of the simplest inversion estimation $\theta_{\mathrm{inv}}$, which depends on the choice of the eigenvalues $\mu_k$~\cite{Wang}. When $\mu_{+}=\mu_{-}$, it is simply given by $\delta\theta
_{\mathrm{bin}}$. For other choices of $\{\mu_k\}$, the performance of $\theta_{\mathrm{inv}}$ cannot outperform that of the MLE and hence the new estimator $\theta _{\mathrm{est}}$, as predicted by the Cr\'{a}mer-Rao inequality~\cite{Helstrom,Braunstein1,Braunstein2,Paris2009}. Finally, in addition to the squeezed-state light interferometry, we believe that our
estimation protocol may also be applicable to other kinds of multi-outcome
measurements (e.g., intensity-difference measurement over the twin-Fock states~\cite{TFs}, which will be shown elsewhere).

\section{Conclusion}

In summary, we have proposed a new data-processing method for the homodyne
detection at one port of squeezed-state light interferometer, where the
measurement quadrature are divided into three bins: $p\in (-\infty ,-a)$, $%
[-a, a]$, and $(a,\infty)$, corresponding to a multi-outcome measurement.
Compared with previous binary-outcome case~\cite{Schafermeier}, we show that
(i) the divergence of phase sensitivity at $\theta=0$ can be removed, which
is useful for estimating a small phase shift; (ii) Higher improvement in the
resolution and the sensitivity is achievable with the realistic experimental
parameters. For instance, we obtain a $38$-fold improvement in the resolution with the average number of photons $\bar{n}\sim427$, while the sensitivity $\sim 1.1e^{-r}/\sqrt{\bar{n}}$, almost approaching the best sensitivity of the single-port homodyne measurement without any data-processing. Furthermore, a new phase-estimation protocol has been developed
based on a combination of the inversion estimators of each outcome. Similar
to the well-known maximum-likelihood estimator, we show that the estimator
is unbiased and its uncertainty can saturate the Cram\'{e}r-Rao bound of
phase sensitivity. Our estimation protocol may also be applicable to other
kinds of multi-outcome measurements.

\begin{acknowledgments}
We thank Professor C. P. Sun for helpful discussions. Project supported by
the National Natural Science Foundation of China (Grant Nos.~91636108,
11775190, 11774024), Science Foundation of Zhejiang Sci-Tech University
(Grant No.~18062145-Y), Open Foundation of Key Laboratory of Optical Field
Manipulation of Zhejiang Province (Grant No.~ZJOFM-2019-002), and the NSFC
program for \textquotedblleft Scientific Research Center" (Grant No.
U1530401).
\end{acknowledgments}

\appendix

\section{Details of Eq.~(3)}

The output state is given by $|\psi _{\mathrm{out}}(\theta )\rangle =\hat{U}%
(\theta )|\psi _{\mathrm{in}}\rangle $, where $\hat{U}(\theta )$ is an
unitary operator%
\begin{eqnarray}
\hat{U}(\theta ) &=&\exp \left( -i\frac{\pi }{2}\hat{J}_{y}\right) \exp
\left( -i\theta \hat{a}^{\dag }\hat{a}\right) \exp \left( -i\frac{\pi }{2}%
\hat{J}_{y}\right)  \notag \\
&=&\exp \left( -i\pi \hat{J}_{y}\right) \exp \left( i\theta \hat{G}\right) ,
\end{eqnarray}%
which represents a sequence actions of the 50:50 beamsplitter at the output
port~\cite{Gerrybook}, the phase accumulation at one of the two paths, and
the 50:50 beamsplitter at the input port. For brevity, we have introduced
Schwinger's representation of the angular momentum $\hat{J}=\frac{1}{2}(\hat{%
a}^{\dag },\hat{b}^{\dag })\hat{\sigma}\binom{\hat{a}}{\hat{b}}$ and $\hat{G}%
=\hat{J}_{x}-\hat{n}/2$, with the Pauli matrix $\hat{\sigma}=(\hat{\sigma}%
_{x},\hat{\sigma}_{y},\hat{\sigma}_{z})$ and $\hat{n}=\hat{a}^{\dag }\hat{a}+%
\hat{b}^{\dag }\hat{b}$.

The ultimate phase-estimation precision is determined by the so-called
quantum Cram\'{e}r-Rao bound~\cite{Helstrom,Braunstein1,Braunstein2,Paris2009}: $\delta \theta _{\mathrm{QCRB}}=1/\sqrt{F_{Q}}$%
, where $F_{Q}$ is the quantum Fisher information. For the unitary
operator $\hat{U}(\theta)$ and the squeezed-state input state $
|\psi _{\mathrm{in}}\rangle =|\alpha _{0}\rangle \otimes |\xi _{0}\rangle$, it
is simply given by%
\begin{equation}
F_{Q}=4\left( \Delta \hat{G}\right) _{\mathrm{in}}^{2}=4\left[ \left\langle
\hat{J}_{x}^{2}\right\rangle _{\mathrm{in}}+\frac{\left( \Delta \hat{n}%
\right) _{\mathrm{in}}^{2}}{4}\right] ,
\end{equation}%
which is optimal when the phases of the two incident light fields satisfy
the phase-matching condition $\cos (\arg \xi _{0}-2\arg \alpha _{0})=-1$,
e.g., $\arg \alpha _{0}=0$ and $\arg \xi _{0}=\pi $~\cite{LPan1,LPan2,LJing}.

For the homodyne detection at one of two ports of the interferometer, the
conditional probability for detecting a measurement quadrature $p\in
(-\infty ,\infty )$ is given by
\begin{equation}
P(p_{a}|\theta )=\int_{-\infty }^{\infty }dx_{a}\int_{-\infty }^{\infty
}dx_{b}\int_{-\infty }^{\infty }dp_{b}W_{\mathrm{out}}(\alpha ,\beta ;\theta
),  \label{Integral}
\end{equation}%
where $\alpha =x_{a}+ip_{a}$ and $\beta =x_{b}+ip_{b}$. The Wigner function
of the output state is given by~\cite{Wang}
\begin{equation}
W_{\mathrm{out}}(\alpha ,\beta ;\theta )=W_{\mathrm{in}}(\tilde{\alpha}%
_{\theta },\tilde{\beta}_{\theta }),  \label{WignerOUT}
\end{equation}%
where
\begin{equation}
\left\{
\begin{array}{l}
\tilde{\alpha}_{\theta }=\alpha \frac{e^{i\theta }-1}{2}+\beta \frac{%
e^{i\theta }+1}{2}, \\
\tilde{\beta}_{\theta }=-\alpha \frac{e^{i\theta }+1}{2}-\beta \frac{%
e^{i\theta }-1}{2}.%
\end{array}%
\right.  \label{ab}
\end{equation}%
Note that Eqs.~(\ref{Integral})-(\ref{ab}) hold for the two-path
interferometer described by $\hat{U}(\theta )$, independent from specific
form of the input state. For the input state $|\alpha_0\rangle%
\otimes|\xi_0\rangle$, we obtain the conditional probabilities for detecting
a field quadrature at one port of the squeezing-state interferometer, as
Eq.~(\ref{pro}) in main text.


\begin{thebibliography}{99}
\bibitem{RMP} L. Pezz\'{e}, A. Smerzi, M. K. Oberthaler, R. Schmied, and P.
Treutlein, ``Quantum metrology with nonclassical states of atomic
ensembles," Rev. Mod. Phys. \textbf{90}, 035005 (2018).

\bibitem{Ma} J. Ma, X. Wang, C. P. Sun, and F. Nori, ``Quantum spin
squeezing," Phys. Rep. \textbf{509}, 89 (2011).

\bibitem{Dowling2008} J. P. Dowling, ``Quantum optical metrology-the lowdown
on high-N00N states," Contemp. Phys. \textbf{49}, 125-143 (2008).

\bibitem{Boto} A. N. Boto, P. Kok, D. S. Abrams, S. L. Braunstein, C. P.
Williams, and J. P. Dowling, ``Quantum Interferometric Optical Lithography:
Exploiting Entanglement to Beat the Diffraction Limit," Phys. Rev. Lett.
\textbf{85}, 2733 (2000).

\bibitem{Giovannetti2004} V. Giovannetti, S. Lloyd, and L. Maccone,
``Quantum-Enhanced Measurements: Beating the Standard Quantum Limit,"
Science \textbf{306}, 1330 (2004).

\bibitem{Durkin2007} G. A. Durkin, and J. P. Dowling, ``Local and Global
Distinguishability in Quantum Interferometry," Phys. Rev. Lett. \textbf{99},
070801 (2007).

\bibitem{Gerry2003} C. C. Gerry, and R. A. Campos, ``Generation of maximally
entangled states of a Bose-Einstein condensate and Heisenberg-limited phase
resolution," Phys. Rev. A \textbf{68}, 025602 (2003).


\bibitem{Braun} D. Braun, G. Adesso, F. Benatti, R. Floreanini, U.
Marzolino, M. W. Mitchell, and S. Pirandola, ``Quantum-enhanced measurements
without entanglement," Rev. Mod. Phys. \textbf{90}, 035006 (2018).

\bibitem{Dorner} U. Dorner, R. Demkowicz-Dobrzanski, B. J. Smith, J. S.
Lundeen, W. Wasilewski, K. Banaszek, and I. A. Walmsley, ``Optimal Quantum
Phase Estimation," Phys. Rev. Lett. \textbf{102}, 040403 (2009).

\bibitem{Zhang2013a} Y. M. Zhang, X. W. Li, W. Yang, and G. R. Jin,
``Quantum Fisher information of entangled coherent states in the presence of
photon loss," Phys. Rev. A \textbf{88}, 043832 (2013).

\bibitem{Resch} K. J. Resch, K. L. Pregnell, R. Prevedel, A. Gilchrist, G.
J. Pryde, J. L. O'Brien, and A. G. White, ``Time-Reversal and
Super-Resolving Phase Measurements," Phys. Rev. Lett. \textbf{98}, 223601
(2007).

\bibitem{Gao} Y. Gao, P. M. Anisimov, C. F. Wildfeuer, J. Luine, H. Lee, and
J. P. Dowling, ``Super-resolution at the shot-noise limit with coherent
states and photon-number-resolving detectors," J. Opt. Soc. Am. B \textbf{27}%
, A170-A174 (2010).

\bibitem{Cohen} L. Cohen, D. Istrati, L. Dovrat, and H. S. Eisenberg,
``Super-resolved phase measurements at the shot noise limit by parity
measurement," Opt. Express \textbf{22}, 11945-11953 (2014).

\bibitem{Distante} E. Distante, M. Je\v{z}ek, and U. L. Andersen,
\textquotedblleft Deterministic Superresolution with Coherent States at the
Shot Noise Limit," Phys. Rev. Lett. \textbf{111}, 033603 (2013).

\bibitem{Schafermeier} C. Schafermeier, M. Jezex, L. S. Madsen, T. Gehring,
and U. L. Andersen, ``Deterministic phase measurements exhibiting
super-sensitivity and super-resolution," Optica \textbf{5}, 60-64 (2018).

\bibitem{LIGO1} The LIGO Scientific Collaboration, ``A gravitational wave observatory operating beyond the quantum shot-noise limit," Nature Phys. \textbf{7}, 962 (2011).

\bibitem{LIGO2} The LIGO Scientific Collaboration, ``Enhanced sensitivity of the LIGO gravitational wave detector by using squeezed states of light," Nature Photon. \textbf{7}, 613 (2013).

\bibitem{Pradyumna} S. T. Pradyumna, E. Losero, I. Ruo-Berchera, P. Traina, M. Zucco, C. S. Jacobsen, U. L. Andersen, I. P. Degiovanni, M Genovese, and T. Gehring, ``Quantum-enhanced correlated interferometry for fundamental physics tests," arXiv:1810.13386[quant-ph].



\bibitem{Feng} X. M. Feng, G. R. Jin, and W. Yang, ``Quantum interferometry
with binary-outcome measurements in the presence of phase diffusion," Phys.
Rev. A. \textbf{90}, 013807 (2014).

\bibitem{Ghirardi} L. Ghirardi, I. Siloi, P. Bordone, F. Troiani, and M. G.
A. Paris, ``Quantum metrology at level anticrossing," Phys. Rev. A \textbf{97%
}, 012120 (2018).

\bibitem{Jin} G. R. Jin, W. Yang, and C. P. Sun, ``Quantum-enhanced
microscopy with binary-outcome photon counting," Phys. Rev. A \textbf{95},
013835 (2017).

\bibitem{YMK} B. Yurke, S. L. McCall, and J. R. Klauder, ``SU(2) and SU(1,1)
interferometers," Phys. Rev. A \textbf{33}, 4033 (1986).

\bibitem{Pezze} L. Pezz\'{e}, A. Smerzi, G. Khoury, J. F. Hodelin, and D.
Bouwmeester, ``Phase Detection at the Quantum Limit with Multiphoton
Mach-Zehnder Interferometry," Phys. Rev. Lett. \textbf{99}, 223602 (2007).



\bibitem{Caves} C. M. Caves, ``Quantum-mechanical noise in an
interferometer," Phys. Rev. D \textbf{23}, 1693 (1981).

\bibitem{Pezze2008} L. Pezz\'{e} and A. Smerzi, ``Mach-Zehnder
Interferometry at the Heisenberg Limit with Coherent and Squeezed-Vacuum
Light," Phys. Rev. Lett. \textbf{100}, 073601 (2008).


\bibitem{LJing} J. Liu, X. Jing, and X. Wang, ``Phase-matching condition for
enhancement of phase sensitivity in quantum metrology," Phys. Rev. A \textbf{%
88}, 042316 (2013).

\bibitem{LPan1} P. Liu, P. Wang, W. Yang, G. R. Jin, and C. P. Sun, ``Fisher
information of a squeezed-state interferometer with a finite photon-number
resolution," Phys. Rev. A \textbf{95}, 023824 (2017).

\bibitem{LPan2} P. Liu, and G. R. Jin, ``Ultimate phase estimation in a
squeezed-state interferometer using photon counters with a finite number
resolution," J. Phys. A: Math. Theor. \textbf{50}, 405303 (2017).

\bibitem{Gerrybook} C. C. Gerry and P. L. Knight, \textit{Introductory
Quantum Optics} (Cambridge University, 2005).


\bibitem{Seshadreesan2} K. P. Seshadreesan, P. M. Anisimov, H. Lee, and J.
P. Dowling, ``Parity detection achieves the Heisenberg limit in
interferometry with coherent mixed with squeezed vacuum light," New J. Phys.
\textbf{13}, 083026 (2011).

\bibitem{Tan} Q. S. Tan, J. Q. Liao, X. G. Wang, and F. Nori, ``Enhanced
interferometry using squeezed thermal states and even or odd states," Phys.
Rev. A \textbf{89}, 053822 (2014).

\bibitem{Wang} J. Z. Wang, Z. Q. Yang, A. X. Chen, W. Yang, and G. R. Jin,
``Multi-outcome homodyne detection in a coherent-state light
interferometer," Opt. Express \textbf{27}, 10343 (2019).


\bibitem{Helstrom} C. W. Helstrom, \textit{Quantum Detection and Estimation
Theory} (Academic, New York, 1976).

\bibitem{Braunstein1} S. L. Braunstein, and C. M. Caves, ``Statistical
distance and the geometry of quantum states," Phys. Rev. Lett. \textbf{72},
3439 (1994).

\bibitem{Braunstein2} S. L. Braunstein, C. M. Caves, and G. J. Milburn,
``Generalized uncertainty relations: Theory, examples, and Lorentz
invariance," Ann. Phys. (NY) \textbf{247}, 135 (1996).

\bibitem{Paris2009} M. G. A. Paris, ``Quantum estimation for quantum
technology," In. J. Quantum Inform. \textbf{7}, 125 (2009).

\bibitem{TFs} M. J. Holland and K. Burnett, ``Interferometric
detection of optical phase shifts at the Heisenberg limit" Phys. Rev. Lett. \textbf{71}, 1355 (1993).

\end{thebibliography}
\end{document}